\documentclass[a4paper,11pt]{article}
\usepackage{jheppub,physics,dsfont,enumitem} 
\usepackage{orcidlink}

\newcommand{\id}{\mathds{1}}
\newcommand{\T}{\scalebox{0.6}{\text{T}}}
\newcommand{\Ord}{\mathcal{O}}
\newcommand{\hak}[1]{\left[ #1 \right]}
\newcommand{\para}[1]{\left( #1 \right)}

\title{Mass spectrum \& linear perturbations of ghost-free multi-spin-2 theory}

\author{J. Flinckman\orcidlink{0009-0004-4545-3123}}
\author{and S. F. Hassan\orcidlink{0000-0003-3910-431X}}
\affiliation{Department of Physics \& The Oskar Klein Centre,\\
Stockholm University, AlbaNova University Centre, SE-106 91 Stockholm, Sweden}

\emailAdd{joakim.flinckman@fysik.su.se}
\emailAdd{fawad@fysik.su.se}

\abstract{
Motivated by the spin-2 nature of gravity, we consider a theory of $\mathcal{N}$ interacting spin-2 fields, formulated in terms of vielbeins, which has been argued to be free of Boulware-Deser ghost instabilities. This is the only known such theory with non-trivial multi-field interactions. We develop an all-order expansion of the vielbein in terms of the metric and Lorentz field perturbations, enabling an expansion of the action in arbitrary backgrounds to any order. The mass matrix is then analysed around proportional vielbein solutions corresponding to vacuum Einstein spacetimes. It is shown that the theory exhibits a non-tachyonic spectrum with one massless and $\mathcal{N}-1$ massive spin-2 modes, regardless of the sign of the cosmological constant. While the massless mode is universal, the non-zero masses cannot generally be determined analytically, but we obtain the lower and upper bounds on each mass eigenvalue in terms of the parameters of the theory. 
}

\arxivnumber{2410.09439}

\begin{document}
\maketitle  

\section{Introduction}
\label{sec:introduction}

The fundamental interactions of nature are formulated in terms of fields characterised by their mass, spin, and charges. These attributes, together with general consistency requirements such as covariance, causality, and absence of ghosts and tachyonic instabilities, more or less uniquely determine the basic structure of the field equations. For fields of spin 0, 1/2, and 1, the resulting theories are the building blocks of the Standard Model of particle physics. An essential feature of the model is the ubiquity of field multiplets, which are not only required for a correct description of nature but also allow for the presence of gauge and global symmetries, which are essential for the theoretical consistency of the model.

In contrast, Einstein's General Relativity is the theory of a single massless spin-2 field, lacking the multiplet structure so crucial for the consistency of the Standard Model. Extending Einstein's theory to consistent theories of multiple spin-2 fields, with or without extra symmetries, has proven challenging. Theoretically, such extensions are of interest because they involve features that are neither known nor fully understood, even at the classical level. Phenomenologically, these models suggest new physics with consequences for gravitational phenomena arising within the spin-2 sector, which is intrinsically linked to gravity.

The main challenge in formulating consistent interactions of spin-2 fields is the presence of ghost instabilities. This paper considers a specific theory of multiple interacting spin-2 fields constructed in \cite{Hassan:2018mcw} and argued to be free of ghosts. This theory, which generalises a subset of the better studied bimetric theories \cite{Hassan:2011zd,Hassan:2018mbl,Hassan:2017ugh} to $\mathcal{N}$ spin-2 fields, is formulated in terms of vielbeins and is, so far, the only known example of such ghost-free interactions. 

Intending to analyse the multivielbein theory, we develop a systematic framework for obtaining an all-order expansion of the vielbein in terms of its corresponding metric and Lorentz field perturbations. We analyse the mass spectrum of quadratic fluctuations by expanding the vielbeins around backgrounds corresponding to Einstein vacuum solutions. In particular, we show that these theories always admit solutions with a tachyon-free spectrum consisting of one massless and ${\mathcal{N}-1}$ massive spin-2 perturbations, with the freedom to choose the cosmological constant independently.

To provide context for the theory considered, we recall that ghosts in non-linear theories of interacting spin-2 fields were identified in the context of massive gravity involving a single massive spin-2 field \cite{Boulware:1972yco}, and the possibility of ghost-free interactions between massless spin-2 fields was ruled out in \cite{Boulanger:2000rq}. However, the development of a fully non-linear action for massive gravity \cite{deRham:2010kj, Hassan:2011vm} and the proof of the absence of ghosts in it \cite{Hassan:2011hr, Hassan:2011tf, Hassan:2011ea}, paved the way for the ghost-free bimetric theory of two interacting spin-2 fields \cite{Hassan:2011zd, Hassan:2011ea, Hassan:2017ugh, Hassan:2018mbl}. The fields, represented by two metrics, interact via a non-derivative potential, $V_{\text{bi}}(g_1,g_2)$, and correspond to one massless and one massive spin-2 mode at the quadratic level. 

Bimetric theory can be trivially extended to multiple spin-2 fields provided they interact only pairwise via bimetric potentials of the type $V_{\text{bi}}(g_1,g_2)+V_{\text{bi}}(g_2,g_3)+V_{\text{bi}}(g_3,g_4)+ \ldots\,$, or $V_{\text{bi}}(g_1,g_2)+V_{\text{bi}}(g_1,g_3)+V_{\text{bi}}(g_1,g_4)+ \ldots\,$, \cite{Khosravi:2011zi, Schmidt-May:2015vnx,Niedermann:2018lhx,Molaee:2019knc,Dokhani:2020jxb,Wood:2024acv,Baldacchino:2016jsz}. The resulting theories propagate one massless spin-2 mode and multiple massive spin-2 modes at the quadratic level \cite{Baldacchino:2016jsz,Wood:2024acv}. However, if the interactions contain loops of bimetric terms, e.g. $V_{\text{bi}}(g_1,g_2)+V_{\text{bi}}(g_2,g_3)+V_{\text{bi}}(g_3,g_1)$, the theory is not ghost-free \cite{Nomura:2012xr, Afshar:2014dta, Scargill:2014wya, deRham:2015cha}. 

Going beyond pairwise couplings, more general interactions between multiple spin-2 fields were considered in \cite{Hinterbichler:2012cn}. That interaction was formulated in terms of antisymmetric products of vielbeins, which also have a limited metric formulation \cite{Hassan:2012wt}. However, such interactions are generically not ghost-free \cite{Scargill:2014wya,Afshar:2014dta,deRham:2015cha}. Therefore, the model constructed in \cite{Hassan:2018mcw} is the only known ghost-free theory of multiple spin-2 fields with multi-field interactions. While it admits certain generalisations, they are not considered here.

The rest of the paper is organised as follows: In Section~\ref{sec:multivielbein_theory}, we review the multivielbein theory of~\cite{Hassan:2018mcw}, presenting the field equations and the Lorentz constraints. Section~\ref{sec:perturbations} describes an all-order expansion of the vielbein perturbations in terms of the metric and Lorentz perturbations, which enables an expansion of the action around generic background solutions to all orders. This is then used to derive the linearised field equations in generic backgrounds. Section~\ref{sec:prop_sol} presents the proportional background solutions to the non-linear field equations in different parametrisations. In Section~\ref{sec:analysis_of_mass_spectrum}, we consider quadratic perturbations around such solutions and analyse the eigenvalues and eigenstates of the mass matrix, in particular, proving the existence of a non-tachyonic spectrum in any vacuum Einstein spacetime. The results are summarised in Section \ref{sec:conclusion}.

\section{The multivielbein theory}
\label{sec:multivielbein_theory}

This section reviews a theory of multiple spin-2 fields interacting via a non-derivative potential term considered in \cite{Hassan:2018mcw} and argued to be free of ghost instabilities that usually plague such interactions. 

In General Relativity the non-linear spin-2 field is described by a metric $g_{\mu \nu}(x)$ which can be decomposed in terms of the vielbein (tetrad) field $e^A_{\;\mu}(x)$ such that $g_{\mu\nu} = e^A_{\;\mu}\,\eta_{AB}\, e^B_{\; \nu}$. Here, we consider a theory of $\mathcal{N}$ interacting spin-2 fields described by the vielbeins $e^A_{I\; \mu }$, $I=1,\ldots,\mathcal{N}$, which so far lacks a formulation in terms of the corresponding metrics $g^I_{\mu \nu}= e^A_{I\; \mu }\eta^{\;}_{AB}e^B_{I\; \nu}$.\footnote{Note that the indices $I,J,\ldots$ are simply labels without any covariant structure and will not be subject to the Einstein summation convention. To simplify the notation, we will write these labels arbitrarily as super- or subscripts wherever there are fewer indices of other types. Otherwise, all Greek letters $\alpha, \beta,...$ are spacetime indices and Latin letters from the beginning of the alphabet $A,B,C,...$ label Lorentz indices.} In the absence of interactions between fields with different values of $I$, the dynamics of each vielbein is governed by an Einstein-Hilbert action. Introducing interactions between the $e^A_{I\; \mu }$ usually gives rise to theories with ghost instabilities. In \cite{Hassan:2018mcw}, it was argued that a theory of $\mathcal{N}$ interacting vielbeins which is ghost-free for any $\mathcal{N}$ is,
\begin{align}
\label{eq:MM_action}
    \mathcal{S} = \int \text{d}^4 x\left[\sum_{I=1}^\mathcal{N} m_I^2 \sqrt{-g_I}\left( R_I - 2 \Lambda_I \right) - V(e_1,\dots, e_{\mathcal{N}})\right] +\sum_{I=1}^\mathcal{N} \mathcal{S}_{\text{M}}^I[e_I, \phi_I]\,. 
\end{align}
The first part of \eqref{eq:MM_action} consists of the Einstein-Hilbert actions for the individual fields with Planck masses $m_I$. The vielbein $e_I$ can minimally couple to matter fields collectively denoted by $\phi_I$ via $\mathcal{S}_{\text{M}}[e_I, \phi_I]$, but for the theory to remain ghost-free, no matter field can interact directly with multiple vielbeins or with fields from other sectors. The interactions between the vielbein fields are contained in the interaction potential given by,
\begin{align}
\label{eq:HSM_interaction}
    V(e_1,\dots, e_{\mathcal{N}}) = 2m^4 \det\Big(\sum_{I=1}^{\mathcal{N}}\beta_I e_I\Big).
\end{align}
Here, a parameter $m$ has been introduced to carry the term's mass dimension, and $\beta_I$ are dimensionless parameters. These, along with the Planck mass parameters $m_I$ and the cosmological constant parameters $\Lambda_I$, constitute the free parameters of the theory. Note that replacing $e_I$ by $\lambda_I e_I$ amounts to the rescalings $\beta_I\to\lambda_I \beta_I$, $m_I\to\lambda_I m_I$, and $\Lambda_I \to \lambda^2_I\Lambda_I$. Hence, one can, for example, absorb the $\beta_I$ into the vielbeins by choosing $\lambda_I=1/\beta_I$, resulting in $m_I \to m_I/\beta_I$ and $\Lambda_I \to \Lambda_I/\beta_I^2$, along with the appropriate scalings of the matter couplings. We will, however, retain all parameters as they appear in the above action. If, for example, we couple the $I=1$ vielbein to the Standard Model matter, then $m_1=M_{\text{Pl}}$ is the physical Planck mass, and the remaining vielbeins are extra spin-2 fields that interact with gravity described by $e_1$.

Before considering the equations of motion, we recall that a vielbein is an invertible matrix with 16 components, compared to the 10 components of the metric. Its additional 6 components are related to the freedom of performing local Lorentz transformations $e^A_{\; \mu}(x) \rightarrow L^A_{\; B}(x) e^B_{\; \mu}(x)$ that leave the metric $g_{\mu \nu}$ invariant, implying that the 6 extra components can be parametrised by a local Lorentz transformation. Since the Einstein-Hilbert action can be formulated entirely in terms of the metric, the Lorentz degrees of freedom of the vielbeins drop out of the derivative part of the action and are retained only in the non-derivative interaction potential \eqref{eq:HSM_interaction}.\footnote{While fermionic matter actions also contain the vielbeins, the local Lorentz invariance of matter couplings ensures that the Lorentz parameters drop out of the matter actions.} Hence, the 6 Lorentz parameters of the vielbein are non-dynamical fields. 

The equations of motion can be derived by varying the action \eqref{eq:MM_action} with respect to the vielbein $e^A_{I\,\mu}$, and give a modified version of the Einstein field equations, which now also include contributions from the interaction potential \eqref{eq:HSM_interaction},
\begin{gather}
\label{eq:det_EoM}
    G^I_{\mu \nu} + \Lambda^{\;}_I g^I_{\mu\nu} + V^I_{\mu \nu} = \frac{1}{2m_I^2}T^I_{\mu \nu}\,, \qquad I = 1,\ldots, \mathcal{N}\,.
\end{gather}
Here $G^I_{\mu \nu}=R^I_{\mu \nu}-\frac{1}{2} g^I_{\mu\nu} R^I$ is the Einstein tensor and $T^I_{\mu \nu}$ is the standard matter energy-momentum tensor given by, $\sqrt{-g_I}\, T^I_{\mu \nu} = -2\delta \mathcal{S}^{I}_{\text{M}}/\delta g_I^{\mu \nu}$. The contribution from the interaction potential can be expressed as,
\begin{gather}
\label{eq:stress_energy}
    V^I_{\mu \nu} = m^4 \frac{\beta_I}{m_I^2}\det(e_I^{-1}u)\,g^I_{\mu \alpha}\, u^{\alpha}_{\;A}\,e^A_{I\; \nu}\,,
\end{gather}
where for convenience, we have defined,
\begin{align}
\label{eq:u}
    u^A_{\; \alpha} = \sum_{I=1}^{\mathcal{N}} \beta_I^{\;} e^A_{I\; \alpha}\,,
\end{align}
and $u^\alpha_{\; A}= (u^{-1})^\alpha_{\; A}$ is the inverse of $u^A_{\; \alpha}$. The existence of the inverse of $u^A_{\;\alpha}$ is guaranteed by the requirement that $\det(u)$ is non-vanishing so that the interaction potential in \eqref{eq:HSM_interaction} is non-trivial. Thus $u^A_{\; \alpha}$ is invertible and can itself be treated like a vielbein such that the associated metric $u^A_{\; \mu}\eta^{\;}_{AB}u^B_{\; \nu}$ produces a non-degenerate null cone.

Note that all terms in \eqref{eq:det_EoM}, except for $V^I_{\mu\nu}$, are symmetric in the two indices. Hence, the antisymmetric part of the equation of motion results in the Lorentz constraints $V^I_{[\mu \nu]} = 0$, which gives, $g^I_{[\mu \alpha}u^{\alpha}_{\;A}e^A_{I\; \nu]}=0$, or equivalently, 
\begin{align}
\label{eq:Lorentz_Constraints_det}
    u^A_{\; [\mu}\eta^{\;}_{AB}e^B_{I \; \nu]}=0\,, \qquad I=1,\ldots\,, \mathcal{N}.
\end{align}
These $6\mathcal{N}$ equations are not all independent. Consider a linear combination obtained by multiplying the left-hand side by $\beta_I$ and summing over $I$. This vanishes on using the definition of $u^{A}_{\; \mu}$ in \eqref{eq:u}, showing that only $6(\mathcal{N}-1)$ of the above equations are linearly independent. It can be shown that the above symmetrisation conditions are the equations of motion for the Lorentz parameters \cite{Hassan:2012wt}. The vielbeins in $V$ contain $6\mathcal{N}$ local Lorentz parameters, but the potential is invariant under local Lorentz transformations that transform all vielbeins in the same way $V(Le_1,\ldots \,,Le_\mathcal{N})= V(e_1,\ldots \,,e_\mathcal{N})$. Thus, $6$ of the Lorentz parameters drop out of $V$, leaving a dependence on $6(\mathcal{N}-1)$ Lorentz fields, consistent with the number of symmetrisation conditions. 

The Lorentz constraints \eqref{eq:Lorentz_Constraints_det} are algebraic conditions that, in principle, can be solved to determine the Lorentz parameters in terms of the metric variables, though such solutions are not known for $\mathcal{N}>2$. For $\mathcal{N}=2$, these reduce to the symmetrisation condition $e^A_{1\; [\mu}\eta^{\;}_{AB}e^B_{2 \; \nu]}=0$ known from the vielbein formulation of bimetric theory \cite{Hinterbichler:2012cn, Zumino:1970tu}. This can be explicitly solved for the Lorentz parameters and allows the theory to be expressed solely in terms of the metrics using the matrix square root $e_1^{-1}e_2^{\;}=\sqrt{g_1^{-1}g^{\;}_2}$ \cite{Deffayet:2012zc,Hinterbichler:2012cn,Hassan:2017ugh}. Such solutions exist only if the null cones of the metrics $g_1$ and $g_2$ overlap~\cite{Hassan:2017ugh}. Thus, for $\mathcal{N}=2$, the potential \eqref{eq:HSM_interaction} gives rise to a subset of the ghost-free bimetric interactions with one fewer parameter. 

In contrast, for $\mathcal{N}>2$, analytic solutions to \eqref{eq:Lorentz_Constraints_det} that express the local Lorentz parameters in terms of the metric variables are yet to be found. Consequently, an equivalent metric formulation of the multivielbein theory is not known.\footnote{However, a restricted approach that retains the Lorentz fields is possible using a method similar to that presented in \cite{Hassan:2012wt} and \cite{Flinckman:2020}.} Nevertheless, the geometric interpretation of the Lorentz constraints remains the same. Namely, to symmetrise the product of vielbeins $e^A_{I\; \mu }$ and $u^A_{\; \mu }$, the null cones of the associated metrics must overlap for the constraints to have real covariant solutions \cite{Hassan:2017ugh}.

The propagating degrees of freedom of the multivielbein action \eqref{eq:MM_action} were analysed in \cite{Hassan:2018mcw} using a Hamiltonian constraint analysis. It was argued that the multivielbein theory propagates $2 +5(\mathcal{N}-1)$ modes at a non-linear order, which is the correct number of degrees of freedom for a theory of one massless and $\mathcal{N}-1$ massive spin-2 fields. In particular, this means that the theory is free of the Boulware-Deser ghosts that would show up as $\mathcal{N}-1$ propagating zero-helicity modes. Thus far, \eqref{eq:MM_action} is the only known ghost-free theory of multiple spin-2 fields with multi-field interactions that go beyond pairwise bimetric couplings of the form $V(e_1,\ldots,e_\mathcal{N})= V_{\text{bi}}(e_1,e_2) + V_{\text{bi}}(e_2,e_3) + \ldots\,$ \cite{Khosravi:2011zi, Nomura:2012xr,  Scargill:2014wya, Schmidt-May:2015vnx, Afshar:2014dta, deRham:2015cha, Baldacchino:2016jsz}.\footnote{See chapter 8.2 in~\cite{Schmidt-May:2015vnx} for a review on the consistent pairwise couplings.} The more general multivielbein interactions considered in~\cite{Hinterbichler:2012cn} are generically not ghost-free \cite{Scargill:2014wya, Afshar:2014dta,deRham:2015cha}.

\section{General perturbations}
\label{sec:perturbations}

The multivielbein theory can be expanded to any order in terms of vielbein perturbations $\delta {e}^A_{I\; \mu}$ around generic background solutions $\overline{e}^A_{I \; \mu}$ to the equations of motion~\eqref{eq:det_EoM}, e.g., cosmological solutions or the proportional backgrounds considered in the next section. Here, we will develop this expansion in terms of the metric perturbations $h^I_{\mu\nu}$ and the perturbations of the Lorentz parameters $w^I_{AB}$. This will be used in Section~\ref{sec:analysis_of_mass_spectrum} to compute the mass spectrum around proportional background solutions. 

\subsection{All-order perturbative expansion}
\label{sec:all_order_perturbations}

To begin, we keep the discussion general and develop an expansion to any order around any given vielbein $\overline{e}^A_{I \; \mu}$. Suppressing the index $I$, let us start with vielbeins $\overline{e}^A_{\;\mu}$ and ${e^A_{\;\mu}=\overline{e}^A_{\;\mu}+\delta e^A_{\;\mu}}$, and their associated metrics $\overline{g}_{\mu\nu}$ and $g_{\mu\nu}=\overline{g}_{\mu\nu}+\epsilon h_{\mu\nu}$, where a small parameter $\epsilon$ is introduced to keep track of the expansions and, later, fix the canonical normalization of the perturbations. We want to express $\delta e^A_{\;\mu}$ in terms of $h_{\mu\nu}$ or, equivalently, ${e}^A_{\;\mu}$ in terms of $g_{\mu\nu}$. This relation must also include the six local Lorentz fields $w_{AB}$. Let us parameterise the vielbein $e^A_{\; \mu}$ as,
\begin{align}
\label{eq:e=LE}
    e^A_{\;\mu} = L^A_{\; B} E^B_{\;\mu} \,,
\end{align}
where $L^A_{\,B}$ is a local Lorentz transformation and $E^B_{\;\mu}$ is a vielbein that is constrained to be fully determined by $g_{\mu\nu}$ and the background quantities. This can be achieved by requiring the product of $E^A_{\,\mu}$ and $\overline{e}^B_{\;\nu}$ to be symmetrised,
\begin{align}
\label{eq:sym-eE}
    \overline{e}^A_{\; \mu}\,\eta_{AB}\, E^B_{\; \nu}=E^A_{\; \mu}\,\eta_{AB}\,\overline{e}^B_{\; \nu}\,.
\end{align}
In matrix notation, the above symmetrisation condition reads $\overline{e}^{\T}\eta\, E=E^{\T}\eta\, \overline{e}$. Since the metrics are ${g=E^{\T}\eta\,E}$ and $\overline{g}=\overline{e}^{\T}\eta\,\overline{e}$, this condition leads to the matrix equation $\overline{g}^{\;-1}g=(\overline{e}^{\;-1}E)^2$ which, on taking the square root, gives the desired result,\footnote{The existence of an acceptable solution $E^A_{\;\mu}$ to the symmetrisation condition~\eqref{eq:sym-eE} is equivalent to the existence of the square-root matrix $\sqrt{\overline{g}^{\;-1}g}$ which is real and transforms as a $(1,1)$ tensor. This requires that the null cones of $g_{\mu\nu}$ and $\overline{g}_{\mu\nu}$ have a non-vanishing overlap~\cite{Hassan:2017ugh}, which is guaranteed if $g_{\mu\nu}$ is a small perturbation of $\overline{g}_{\mu\nu}$. Otherwise, the results are also valid for large $\epsilon h_{\mu\nu}$ when restricted to produce overlapping null cones.}
\begin{align}
\label{eq:E=eSQRT}
    E^B_{\;\mu}=\overline{e}^B_{\;\nu}\,\left(\sqrt{\overline{g}^{\;-1}g}\right)^\nu_{\;\mu}\,.
\end{align}
In terms of the metric perturbation we have $(\overline{g}^{-1}g)^\mu_{\;\nu}=\delta^\mu_{\;\nu}+ \epsilon h^\mu_{\;\nu}$, where $h^\mu_{\;\nu}=\overline{g}^{\mu\lambda} h_{\lambda\nu}$. Now, the matrix $\sqrt{\id+\epsilon h}$ can be expanded in powers of the matrix $h$ with components $ h^\mu_{\; \nu}$. Then~\eqref{eq:e=LE} becomes,   
\begin{align}
\label{eq:e=eLh}
    e^A_{\; \mu}=\overline{e}^A_{\; \nu}\,L^\nu_{\;\lambda}\,\sum_{n=0}^{\infty}\binom{1/2}{n}{\epsilon^n(h^n)^\lambda_{\; \mu}}\;, \quad \text{where,} \quad \binom{1/2}{n}=\left \lbrace
    \begin{array}{cc}
        1 & ~~n=0  \\
        \frac{(-1)^{n-1}}{2^{2n-1}n}\binom{2n-2}{n-1} & ~~n>0
    \end{array}
     \right. \;.
\end{align}
Here, for later convenience, we have introduced $L^\nu_{\;\lambda}=\overline{e}^\nu_{\; A}\,L^A_{\; B}\,\overline{e}^B_{\;\lambda}$ which is a $\overline{g}$-congruence, i.e. $L^\mu_{\; \alpha}\,\overline{g}_{\mu \nu}L^\nu_{\; \beta}=\overline{g}_{\alpha \beta}$.

The Lorentz transformation $L^A_{\; B}$ can be parameterised in terms of the six independent fields $w^{\;}_{AB}=-w^{\;}_{BA}$ using the Cayley transform. Introducing a matrix $w$ with elements $w^A_{\;B}=
\eta^{AC}\,w^{\;}_{CB}$, the Cayley transform of $L^A_{\; B}$ is,
\begin{align}
\label{eq:Cayley}
    L^A_{\; B} = \big[(\id+ w)^{-1}\big]^A_{\;C}\big[\id-w\big]^C_{\;B} 
    \quad\implies\quad
    L^\mu_{\; \nu}=\big[(\id+\omega)^{-1}\big]^\mu_{\;\alpha}\big[\id-\omega\big]^\alpha_{\;\nu}\,.
\end{align}
In the second equation, $\omega$ is a matrix with elements $\omega^\mu_{\;\nu}=\overline{e}^\mu_{\;A}\, w^A_{\;B}\,\overline{e}^B_{\;\nu}$. From~\eqref{eq:e=LE} and~\eqref{eq:E=eSQRT} it follows that when $g_{\mu\nu}$ and $e^A_{\;\mu}$ are restricted to their background values, i.e. $g_{\mu \nu}=\overline{g}_{\mu \nu}$ and $e^A_{\; \alpha}= \overline{e}^A_{\; \alpha}$, then $\overline{L}^A_{\;B}=\delta^A_{\;B}$ and $\overline{\omega}^\mu_{\; \nu}=0$ at the background level.\footnote{A non-trivial $\overline{L}$ will result if, instead of~\eqref{eq:sym-eE}, $E$ is symmetrised with a Lorentz transform of $\overline e$.} Hence, $\omega^\mu_{\;\nu}$ are small perturbations around this background and $L^\mu_{\;\nu}$ can be expanded as,
\begin{align}
\label{eq:L-expansion}
    L^\mu_{\;\nu}=\delta^\mu_\nu + 2 \sum_{n=1}^\infty (-1)^n \epsilon^n(\omega^n)^\mu_{\; \nu}\,.
\end{align}
The perturbative parameter $\epsilon$ is again introduced to keep track of the expansion. Finally, substituting this in~\eqref{eq:e=eLh} yields an all-order expansion of the vielbein in powers of the metric and Lorentz perturbations,
\begin{align}
\label{eq:all_ord_exp}
    e^A_{\; \mu} 
    = \overline{e}^A_{\;\mu} + \overline{e}^A_{\; \nu}\sum_{n=1}^\infty \epsilon^n \hak{\binom{1/2}{n}(h^n)^\nu_{\; \mu}+2 \sum_{m=1}^n (-1)^m \binom{1/2}{n-m}(\omega^m)^\nu_{\; \lambda}(h^{n-m})^\lambda_{\; \mu} }.
\end{align}
From this, one can read off the perturbations of all vielbeins $\delta e^A_{I\;\mu}=e^A_{I\;\mu}- \overline{e}^A_{I\;\mu}$ in terms of the corresponding metric and Lorentz perturbations to any order in $\epsilon$.

Up to this point, the expansion has been completely general and can be applied to any vielbein theory. Using this for each of the vielbeins, we can now obtain an all-order expansion of the multivielbein potential~\eqref{eq:HSM_interaction}. Note that, 
\begin{gather}
\label{eq:pot_expans}
    V=-2m^4\det(\sum_{I=1}^{\mathcal{N}} \beta_I e_I)
    =-2m^4\det(\sum_{I=1}^{\mathcal{N}} \beta_I (\overline{e}_I+\delta e_I)) 
    =\overline{V}\det\Big(\id+\Delta u\Big)\,, 
\end{gather}  
where $\overline{V}=-2m^4 \det(\overline{u})$ with  $\overline{u}^A_{\; \mu}= \sum_I \beta_I \overline{e}^A_{\;\mu}$, and, 
\begin{gather}
\label{eq:Delta-u}
    (\Delta u)^\mu_{\;\nu}=\sum_{I=1}^{\mathcal{N}}\beta_I\,\overline{u}^\mu_{\;A}\delta e^A_{I\;\nu}\,.
\end{gather}  
Here, $\delta e^A_{I \; \mu}=e^A_{I \; \mu}-\overline{e}^A_{I \; \mu}$ can be read off from~\eqref{eq:all_ord_exp}. The determinant can be expressed as $\det(\id+\Delta u)=\sum_{n=0}^4 \sigma_n(\Delta u)$, where $\sigma_n(\Delta u)$ are the elementary symmetric polynomials of the eigenvalues associated with the matrix $\Delta u$ and are given explicitly by,
\begin{align}
\label{eq:el_sym_pol}
    &\sigma_0(\Delta u)=1\,, \qquad \sigma_1(\Delta u)=\Tr (\Delta u)\,, 
    \qquad \sigma_2(\Delta u)=\frac{1}{2}\hak{\Tr^2(\Delta u)-\Tr(\Delta u^2)}\,,\\
    &\sigma_3(\Delta u)=\frac{1}{6}\hak{\Tr^3(\Delta u)-3\Tr(\Delta u)\Tr(\Delta u^2)+2\Tr(\Delta u)^3}\!, \qquad \sigma_4(\Delta u)=\det(\Delta u)\,.\notag
\end{align}
The expansion terminates since for $4\times 4$ matrices, $\sigma_{n>4}(\Delta u)=0$. With $\Delta u$ determined by~\eqref{eq:Delta-u} and~\eqref{eq:all_ord_exp}, we have an all-order expansion of the interaction potential in powers of $h^\mu_{I\;\nu}$ and $\omega^\mu_{I\;\nu}$ around generic backgrounds. Note that the Lorentz perturbations appear algebraically in the equations of motion and can, in principle, be eliminated from the action in favour of the metric perturbations by solving the Lorentz constraints~\eqref{eq:Lorentz_Constraints_det}.

For completeness, recall that the Einstein-Hilbert action also has a reasonably straightforward all-order expansion in terms of the metric perturbations $g_{\mu\nu}=\overline{g}_{\mu\nu}+ \epsilon h_{\mu\nu}$. This follows from the fact that the Ricci curvatures of $g_{\mu\nu}$ and $\overline{g}_{\mu\nu}$ are related by,
\begin{align}
\label{eq:Ricci_expansion}
    R_{\alpha \beta} &= \overline{R}_{\alpha \beta} + 2\epsilon\overline{\nabla}^{\;}_{\left[\mu \right. }C^{\mu}_{\;\; \left.\alpha\right] \beta} +  2\epsilon^2C^{\mu}_{\;\; \sigma [\mu }C^{\sigma}_{\;\;  \alpha] \beta}\,,\\
    C^{\mu}_{\;\; \alpha \beta}
    &=    \frac{1}{2}g^{\mu \nu}(\overline{\nabla}_{\alpha} h_{\nu \beta}+\overline{\nabla}_{\beta} h_{\alpha\nu}-\overline{\nabla}_{\nu} h_{\alpha \beta})\,.
\end{align}
Hence, the derivative terms contribute at most two powers of $\epsilon h_{\mu \nu}$. All higher-order contributions result from the binomial expansion of $g^{\mu\nu}=[(\id+\epsilon h)^{-1}]^\mu_{\;\rho} \,\overline{g}^{\rho\nu}$ and the expansion,
\begin{align}
    \sqrt{-g}=\sqrt{-\det(g)}=\sqrt{-\det(\overline{g})}\sqrt{\det(\id +\epsilon h)}.   
\end{align}
The formalism described above applies not only to the action~\eqref{eq:MM_action} but allows any vielbein interaction to be expanded to all orders in the metric and Lorentz perturbations. While we have described the expansion of the action, below we will use ~\eqref{eq:all_ord_exp} to expand the equation of motion.

\subsection{Linearised field equations}
\label{sec:linear_field_equations}

As an application of the expansion, we proceed to derive the linearised field equations, which can be used to study the evolution of perturbations. Expanding the equations of motion~\eqref{eq:det_EoM} to first order in $\epsilon_I$, around vielbeins $\overline{e}^A_{I \; \mu }$ satisfying the background equations ${\overline{G}^I_{\mu \nu} + \Lambda_I \overline{g}^I_{\mu\nu} + \overline{V}^I_{\mu \nu} = \tfrac{1}{2m_I^2}\overline{T}^I_{\mu \nu}}$, yields the linearised field equations,
\begin{align}
\label{eq:linear_field_equations}
    \epsilon_I\mathcal{E}_{\mu \nu}^{I  \;\; \alpha \beta}h^I_{\alpha \beta} +\tfrac{1}{2}\epsilon_I
    \para{\delta^\alpha_\mu \delta^\beta_\nu(\overline{R}_I-2\Lambda_I)-\overline{g}_{\mu \nu}\overline{R}_I^{\alpha \beta}}h^I_{\alpha \beta}- \delta V^I_{\mu \nu}=- \frac{1}{2m_I^2} \delta T^I_{\mu \nu} + \Ord(\epsilon_I^2)\,,
\end{align}
where $\delta T^I_{\mu \nu}=T^I_{\mu \nu}-\overline{T}^I_{\mu \nu} + \Ord(\epsilon_I^2)$ are matter perturbations, $\delta V^I_{\mu \nu} = V^I_{\mu \nu}- \overline{V}^I_{\mu \nu}+ \Ord(\epsilon_I^2)$ are perturbations of the multivielbein interaction and $\mathcal{E}_{\mu \nu}^{I  \;\; \alpha \beta}$ is the differential operator defined in terms of the $\overline{g}_{\mu \nu}$-compatible covariant derivative,
\begin{align}
    \label{eq:Lichnerowicz}
    \mathcal{E}_{\mu \nu}^{\;\;\;\alpha \beta}h_{\alpha \beta} = \frac{1}{2}\left(\delta^\alpha_\mu \delta^\beta_\nu \overline{\nabla}^2- \overline{g}_{\mu \nu}\overline{g}^{\alpha \beta}\overline{\nabla}^2 +\overline{g}^{\alpha \beta}\overline{\nabla}_\mu \overline{\nabla}_\nu + \overline{g}_{\mu \nu}\overline{\nabla}^\alpha \overline{\nabla}^\beta- 2\delta^{\alpha}_{(\mu}\overline{\nabla}^\beta\overline{\nabla}_{\nu)}\right)h_{\alpha \beta}\,.
\end{align}
We now proceed to compute $\delta V^I_{\mu \nu}$. From~\eqref{eq:all_ord_exp}, one obtains the expansion of the vielbeins and their inverses to first order in $\epsilon_I$,
\begin{align}
\label{eq:gen_vielbein_expansion}
    e^A_{I \; \mu} &= \overline{e}^A_{ I\; \mu}+\frac{\epsilon_I}{2}\overline{e}^A_{ I\; \nu}\para{h^\nu_{I \; \mu}-4\omega^\nu_{I \; \mu}}+ \Ord(\epsilon_I^2)\,,\qquad  \\
    e^\mu_{I\; A} &= \overline{e}^\mu_{I \;A}- \frac{\epsilon_I}{2}\para{h^\mu_{I \; \nu}-4\omega^\mu_{I \; \nu}}\overline{e}^\nu_{I \; A} + \Ord(\epsilon_I^2)\,.
\end{align}
To compute $\delta V^I_{\mu \nu}$, it is convenient to define a variable $X^\alpha_{I\;\beta}$ that relates $(e^{-1}_Iu)^\alpha_{\;\beta}$ and the corresponding background expression $(\overline{e}_I^{-1}\overline{u})^\alpha_{\;\beta}$ such that,
\begin{gather}
\label{eq:vielbein_product}
    (e^{-1}_{I }u)^{\alpha}_{\; \mu} = \para{\delta^\alpha_\beta +  X^\alpha_{I\; \beta}}(\overline{e}_I^{-1}\overline{u})^\beta_{\;\mu}+ \Ord(\epsilon^2)\,.
\end{gather}
Using the expansions for the vielbeins and \eqref{eq:u}, one obtains,
\begin{gather}
\label{eq:X_I_def}
    X^\alpha_{I \; \beta}=\overline{e}^\alpha_{I \; A}\sum_J  \frac{\epsilon_J\beta_J}{2} \overline{e}^A_{J \; \mu}\para{h^\mu_{J \; \nu}-4\omega^\mu_{J \; \nu}}\overline{u}^\nu_{\;B}\overline{e}^B_{I \; \beta}-\frac{\epsilon_I}{2}\para{h^\alpha_{I \; \beta}-4\omega^\alpha_{I \; \beta}}.
\end{gather}
Substituting~\eqref{eq:vielbein_product} into the expression for $V_{\mu\nu}$ and using $\overline{V}^\mu_{I \; \nu}=\beta_I \frac{m^4}{m_I^2}\det(\overline{e}^{-1}_I \overline{u})\overline{u}^\mu_{\; A}\overline{e}^A_{I \; \nu}$, the potential term takes the form, 
\begin{align}
    V^I_{ \mu \nu}
    &= g^I_{\mu \alpha}\overline{V}^\alpha_{I \; \beta}\det(\id +   X_I)\hak{(\id +  X_I)^{-1}}^\beta_{\; \nu}+ \Ord(\epsilon^2)\notag\\
    &= \overline{V}^I_{\mu \nu}- \hak{ \overline{V}^I_{\mu \beta}\para{X^\beta_{I \; \nu}-\delta^\beta_{\nu}X^\sigma_{I \; \sigma}} - \epsilon_I h^I_{\mu \alpha}\overline{V}^\alpha_{I \; \nu}} + \Ord(\epsilon^2)\,,
\end{align}
where we have used $\det(\id+X_I)\big(\id+X_I\big)^{-1}=\id-\big(X_I-\id\Tr(X_I) \big)+ \Ord(X_I^{2})$. $\delta V^I_{\mu \nu}$ can now be identified and using~\eqref{eq:X_I_def}, takes the form,
\begin{align}
    \delta V^I_{\mu \nu} &= \overline{V}^I_{\mu \beta}\para{X^\beta_{I \; \nu}-\delta^\beta_{\nu}X^\sigma_{I \; \sigma}} - \epsilon_I h^I_{\mu \alpha}\overline{V}^\alpha_{I \; \nu}\notag\\
    &= \overline{V}^I_{\mu \alpha}\hak{\overline{e}^\alpha_{I \; A}\sum_J \frac{\epsilon_J\beta_J}{2} \overline{e}^A_{J \; \beta}\para{h^\beta_{J \; \gamma}-4\omega^{\beta}_{J \; \gamma} }\overline{u}^\gamma_{\; B}\overline{e}^B_{I \; \nu}}- \frac{\epsilon_I}{2}\overline{V}^I_{\mu \alpha}\Big(h^\alpha_{I \; \nu}-4\omega^\alpha_{I \; \nu} \Big)\notag\\
\label{eq:spin2_stress_pert}
    &-\epsilon_Ih^I_{\mu \alpha}\overline{V}^\alpha_{I\; \nu }
    -\overline{V}^I_{\mu \nu}\hak{\sum_J \frac{\epsilon_J\beta_J}{2}\Tr(\overline{u}^{-1}\overline{e}_Jh_J)- \frac{\epsilon_I}{2}\Tr(h_I)}.
\end{align}
Substituting this into~\eqref{eq:linear_field_equations} gives an explicit form of the linear field equations. 

These equations include the linearised Lorentz constraints, $\delta V^I_{[\mu \nu]}=0$. Since the background equations, $\overline{V}_{[\mu \nu]}^I=0$ are assumed to be satisfied, the last term  in~\eqref{eq:spin2_stress_pert} does not contribute, and the linearised Lorentz constraints can be put in the form,
\begin{align}
\label{eq:lin_LC}
    \sum_J \beta_J \hak{\epsilon_J\overline{e}^A_{I \; \mu}\overline{e}^\alpha_{J \; A}\para{h^J_{\alpha \nu}-4\omega^J_{\alpha \nu} }+\epsilon_I\para{h^I_{\mu \alpha} +4\omega^I_{\mu \alpha} }\overline{e}^\alpha_{I \; A} \overline{e}^A_{J \; \nu} }= \Big[ \mu \leftrightarrow \nu \Big],
\end{align}
consistent with the linearisation of~\eqref{eq:Lorentz_Constraints_det}. These equations determine the $\mathcal{N}-1$ non-dynamical Lorentz parameters $\omega^I_{\alpha \beta}$ yielding field equations dependent only on the metric perturbations $h^I_{\mu \nu}$. Note that $\omega^I_{\alpha\beta}=0$ is not a solution generically, hence, the Lorentz parameters may be non-trivial even at the lowest order, depending on the background solution.

\section{Proportional background solutions}
\label{sec:prop_sol}

This section examines proportional solutions to the multivielbein field equations~\eqref{eq:det_EoM} which are interesting since they coincide with the vacuum solutions of General Relativity with a cosmological constant. We will see in Section~\ref{sec:analysis_of_mass_spectrum} that perturbations around such solutions exhibit Fierz-Pauli mass terms, which is not the case for more general solutions. Hence, proportional solutions are useful for obtaining the mass spectrum of the multivielbein theory. In this section, we show that proportional solutions always exist and we will provide explicit parametric solutions to the consistency conditions that arise from the field equations.

\subsection{Proportional backgrounds}
\label{sec:proportional_backgrounds}

An important class of solutions to the field equations~\eqref{eq:det_EoM} with $T^I_{\mu\nu}=0$ is given by proportional vielbeins. Let us start from an Ansatz of the form $e^A_{I\; \mu}(x) = c_I(x) \overline{e}^A_{\; \mu}(x)$, where $c_I$ are non-zero real scalar functions and $\overline{e}^A_{\;\mu}$ is a common background vielbein.\footnote{In fact, it is enough to start with the more general Ansatz of proportional metrics, $g^I_{\mu\nu}=c_I^2\,\overline{g}_{\mu\nu}$, which for the vielbeins would imply $e^A_{I\; \mu} = c_IL^A_{I\; B}\overline{e}^B_{\; \mu}$ where $L^A_{I \; B}$ is a local Lorentz transformation. Then the Lorentz constraints~\eqref{eq:Lorentz_Constraints_det} imply that all $L^A_{I\; B}$ are the same, leading to the proportional vielbeins Ansatz.} The proportional Ansatz inserted into the field equations~\eqref{eq:det_EoM} yields,
\begin{align}
\label{eq:background_solution}
    \overline{G}_{\mu \nu}+ \para{ \Lambda_I c_I^2 + m^4k^3 
    \frac{\beta_I}{m_I^2 c_I}}\overline{g}_{\mu \nu}=0\,, \qquad I=1,\ldots, \mathcal{N},
\end{align}
where we have defined $k = \sum_I \beta^{\;}_I c_I^{\;}$, and $\overline{G}_{\mu \nu}$ is the Einstein tensor corresponding to the background metric $\overline{g}_{\mu \nu}=\overline{e}^A_{\; \mu}\eta^{\;}_{AB} \overline{e}^B_{\; \nu}$. Taking the covariant derivative of~\eqref{eq:background_solution} yields $\partial_{\mu}\left(\Lambda_I c_I^2 + {m^4k^3}\frac{\beta_I}{m_I^2 c_I} \right) = 0$, hence the expression in the parentheses must be a constant for each $I$. This provides $\mathcal{N}$ equations that determine the $c_I(x)$ in terms of the constants in these equations, therefore, the $c_I$ are themselves constants, $\partial_\mu c_I=0$. Consequently,~\eqref{eq:background_solution} reduces to the Einstein equation for $\overline{g}_{\mu \nu}$ with a cosmological constant,
\begin{align}
\label{eq:CC}
    \Lambda=\Lambda_I c_I^2 +  m^4k^3 \frac{\beta_I}{m_I^2 c^{\;}_I}\,, \qquad  \; I=1,\ldots,\mathcal{N}.
\end{align}
For the $\mathcal{N}$ equations~\eqref{eq:background_solution} to be compatible, the cosmological constant $\Lambda$ must be the same for all $I$, yielding the consistency conditions,
\begin{align}
\label{eq:c_consistency}
    \Lambda_1c_1^2 +m^4k^3 \frac{\beta_1}{m_1^2c_1^{\;}}=
    \Lambda_2c_2^2 +m^4k^3 \frac{\beta_2}{m_2^2c_2^{\;}}=
    \ldots=\Lambda_\mathcal{N}c_\mathcal{N}^2 +m^4k^3 
    \frac{\beta_\mathcal{N}}{m_\mathcal{N}^2 c_\mathcal{N}^{\;}}\,.
\end{align}
Note that in the above equations, the $c_I$ can always be understood to appear in the combinations $\Lambda^{\;}_I c_I^2$, $\beta_I c_I$, and $m_I c_I$, which is a reflection of the scaling property discussed below equation \eqref{eq:HSM_interaction}. We can take equations \eqref{eq:c_consistency} to determine the $c_I$ in terms of the free parameters $\beta_I, m_I$, and $\Lambda_I$ that appear in the multivielbein action. We will show below that such solutions always exist with the desired property of leading to non-tachyonic masses for spin-2 perturbations propagating on the background spacetime. The solution then determines the effective cosmological constant $\Lambda$ in terms of the free parameters, such that \eqref{eq:background_solution} reduce to Einstein's vacuum equation with a cosmological constant $\Lambda$,  hence the solution space of the multivielbein theory contains all Einstein vacuum solutions. 

Alternatively, for phenomenological applications, it will be convenient to regard $\Lambda$ and the $c_I$ as free parameters (say, to be determined through observations) and use equations~\eqref{eq:c_consistency} to parameterise a subset of the theory's parameters $(\beta_I, m_I,\Lambda_I)$ in terms of the $c_I$. We will later construct such solutions explicitly.

\subsection{Proof of existence of solutions}
\label{sec:existence_of_solutions}
To prove that the consistency conditions always have solutions for the $c_I$, we first note that~\eqref{eq:c_consistency} is scale-invariant in the sense that under the scaling $c_I \mapsto \lambda c_I$, for all $I$, all terms scale by a factor of $\lambda^2$ which cancels. Hence, we can, without loss of generality, scale all $c_I$ such that $c_1=\beta_1^{-1}$ and define $y^{\;}_I=\beta_I c_I$ for the remaining $I=2,\ldots\,, \mathcal{N}$. Taking the difference between the $I^{\text{th}}$ and the $I=1$ expression in \eqref{eq:c_consistency}, moving all terms to one side, and multiplying by $y^{\;}_I$, we define the polynomials,
\begin{align}
\label{eq:consistency_polynomial}
    P_I(y_2^{\;},\ldots,y_{\mathcal{N}}^{\;}) = m^4 \para{1+\,\sum_{J=2}^\mathcal{N}  y^{\;}_J}^3 \para{\frac{\beta_1^2}{m_1^2}y^{\;}_I-\frac{\beta_I^2}{m_I^2}}+y^{\;}_I\para{\frac{\Lambda_1}{\beta_1^2}-\frac{\Lambda_I}{\beta_I^2}y_I^2}.
\end{align}
The consistency conditions are now equivalent to the system of $\mathcal{N}-1$ coupled fourth-order polynomial equations $P_I=0$, in $\mathcal{N}-1$ variables $y^{\;}_I$. Such systems of equations may have no solutions at all, or they may, in principle, have infinitely many solutions.

We now prove that this system of equations always has at least one real solution in the first quadrant, i.e. where all $y^{\;}_I>0$. If $y^{\;}_I=0$, the $I^{\text{th}}$ polynomial reduces to,
\begin{align}
    P_I(y^{\;}_2,\ldots,y^{\;}_I=0,\ldots\,,y^{\;}_\mathcal{N})=-m^4\frac{\beta_I^2}{m_I^2}\para{1+\sum_{J=2}^\mathcal{N} y^{\;}_J}^3 < 0\,, 
\end{align}
which is strictly negative if $y^{\;}_J > 0 $ for $J\neq I$. On the other hand, if $y^{\;}_I$ is very large and positive, its dominant $4^{\text{th}}$-order contribution is contained in,
\begin{align}
    P_I(y_2^{\;},\ldots\,,y_{\mathcal{N}}^{\;}) = m^4\frac{\beta_1^2}{m_1^2}\para{1+\sum_{J=2}^\mathcal{N} y^{\;}_J}^3y^{\;}_I+\ldots > 0\,,
\end{align}
which is positive when $y^{\;}_J> 0$. This means that $P^{\;}_I$ change sign on the positive $y^{\;}_I$-axis, implying that there exists at least one continuous surface in the first quadrant where $P_I=0$. Since this is true for all $P_I$, it follows from the Poincaré-Miranda theorem \cite{Miranda:1940, Kulpa:1997} that these surfaces must all intersect in at least one point.\footnote{Poincaré-Miranda theorem: If $\mathcal{N}$ continuous functions of $\mathcal{N}$ variables ${P_I: [a_1, b_1]  \times \ldots \times [a_\mathcal{N}, b_\mathcal{N}] \rightarrow \mathbb{R} }$ satisfy the inequalities $ P_I(y_1, \ldots, y_{I-1}, a_I, y_{I+1}, \ldots, y^{\;}_\mathcal{N}) \leq 0 $ and $ P_I(y_1^{\;}, \ldots, y_{I-1}, b_I, y_{I+1}, \ldots, y^{\;}_\mathcal{N}) \geq 0 $ for all $y_{J\neq I}$ in the domain and all $I$, then there exists a point $(x_1,\ldots,x_\mathcal{N})$ in the domain where $ P_I(x_1,\ldots,x_\mathcal{N}) = 0 $ for all $P_I$ simultaneously.} The consistency equations~\eqref{eq:c_consistency} thus always have a solution such that $y^{\;}_I=\beta_I c_I>0$. 

More than one such solution may exist in the first quadrant, and the equations~\eqref{eq:consistency_polynomial} generically also have solutions elsewhere. However, we will see in the next section that the condition $\beta_Ic_I >0$ is enough to guarantee that the spin-2 perturbations are all non-tachyonic. 

We make a few observations before proceeding further:
\begin{enumerate}[label=(\roman*)]
    \item If all bare cosmological constant parameters vanish, $\Lambda_I=0$, then equations \eqref{eq:c_consistency} admit the solution $c_I=\lambda \beta_I/m_I^2$ with a scale parameter $\lambda$. This leads to a positive effective cosmological constant $\Lambda=m^4\lambda^2(\sum_I \beta_I^2/m_I^2 )^3 > 0$, and as will be shown later, the non-zero spin-2 masses will then be degenerate and equal to $\Lambda$.
    \item If $\Lambda=0$ then $\Lambda_I c_I^2 =-m^4k^3\frac{\beta_I c_I}{m_I^2 c^{2}_I}$ for all $I$. Hence, we need $\Lambda_I\neq 0$ for flat spacetime solutions to exist, and the conditions $\beta_I c_I>0$ imply $\Lambda_I<0$, which we will later show prohibits tachyonic masses.
    \item The cosmological constant $\Lambda$ identified through \eqref{eq:background_solution} is associated with the background metric $\overline{g}_{\mu\nu}$. The cosmological constants associated with the metrics $g_I= c_I^2\;\overline{g}$ are, however, $\Lambda/c_I^2$. Hence, if we identify $g_1$ as the physical metric that would couple to standard model fields, then $\Lambda/c_1^2$ will be the corresponding physical cosmological constant. 
\end{enumerate}

\subsection{Parametric solutions}
\label{sec:parametric_sol}

We have demonstrated that there exist proportionality constants $c_I$ satisfying the consistency conditions~\eqref{eq:c_consistency} for all parameter values. This, in turn, determines the effective cosmological constant in~\eqref{eq:background_solution} in terms of the free parameters of the action, although, in practice, finding explicit solutions for $c_I$ is non-trivial. 

Alternatively, the consistency conditions~\eqref{eq:c_consistency} can be solved explicitly for any of the three sets of free parameters $\beta_I, m_I$ or $\Lambda_I$. For example, solving for the $\beta_I$ amounts to parametrising them in terms of the $c_I$ and the effective cosmological constant $\Lambda$ which then replace the $\beta_I$ as free parameters.\footnote{Note that due to the invariance of \eqref{eq:c_consistency} under the scaling $\lambda c_I, \lambda^2\Lambda$, the $c_I$ and $\Lambda$ collectively constitute only $\mathcal{N}$ free parameters.} The advantage of this approach is that explicit solutions can be obtained, and the physical cosmological constant $\Lambda$ \eqref{eq:CC}, appears as a free parameter in the theory. Regarding $\Lambda$ as a independent parameter, the consistency conditions \eqref{eq:c_consistency}, take the same form a \eqref{eq:CC},
\begin{align}
\label{eq:cosmological_constant}
    \Lambda=\Lambda_I c_I^2 +  m^4k^3 \frac{\beta_I}{m_I^2 c^{\;}_I}\,, \qquad  \; I=1,\ldots,\mathcal{N}.
\end{align}
Without loss of generality, we can solve the $I=1$ equation for $k^3=\frac{m_1^2 c^{\;}_1}{m^4\beta_1}(\Lambda -\Lambda^{\;}_1 c_1^2)$. Substituted into the remaining equations, the $\beta_I$ parameters can explicitly be solved for, yielding,
\begin{align}
\label{eq:c_I_sol}
    \beta_I = \beta_1\frac{m_I^2c_I^{\;}}{m_1^2 c_1^{\;}}\frac{\Lambda-\Lambda^{\;}_Ic_I^2}{\Lambda-\Lambda^{\;}_1c_1^2}\,, \qquad I = 2,\ldots,\mathcal{N}.
\end{align}
These solutions and the form of $k^3$ must be consistent with the definition of $k=\sum_I\beta_Ic_I$. Inserting~\eqref{eq:c_I_sol} into $k^3 = (\sum_I \beta_I c_I)^3$ one can solve for $\beta_1$,
\begin{align}
\label{eq:beta_1_sol}
    \beta_1^4 = \frac{1}{m^4 c_1^4}\frac{\hak{m_1^2c_1^2(\Lambda-\Lambda^{\;}_1c_1^2)}^4}{\hak{\sum_{I=1}^\mathcal{N}m_I^2c_I^2(\Lambda-\Lambda^{\;}_Ic_I^2)}^3}\;.
\end{align}
Taking the fourth root, the solution for $\beta_1$ will be real if the denominator is positive. We will later see that we want to impose $\Lambda>\Lambda_Ic_I^2$ to guarantee that the spin-2 perturbations are non-tachyonic. 

Substituting $\beta_1$ into~\eqref{eq:c_I_sol}, we can write all the $\beta_I$ in the closed form,
\begin{align}
\label{eq:beta_I_sol}
    \beta_I = \frac{m_I^2c^{\;}_I}{m}\frac{\Lambda-\Lambda^{\;}_Ic_I^2}{\hak{\sum_{J=1}^\mathcal{N} m_J^2c_J^2(\Lambda-\Lambda^{\;}_Jc_J^2) }^{3/4}}\;, \qquad I=1,\ldots,\mathcal{N}.
\end{align}
From which the useful relation $k=\sum_I \beta_I c_I=\frac{1}{m}\hak{\sum_I m_I^2c_I^2(\Lambda-\Lambda^{\;}_Ic_I^2)}^{1/4}$ can be obtained. Note that the condition $\beta_I c_I>0$ seen in the previous section implies $\Lambda-\Lambda_Ic_I^2>0$, therefore, the real parametric solutions lie in the first quadrant. For solutions with a positive cosmological constant $\Lambda >0$, the condition $\Lambda -\Lambda^{\;}_Ic^2_I>0$ allows both positive and negative $\Lambda_I$. However, for $\Lambda\leq 0$, only negative $\Lambda_I$ are admissible.  

It can be verified that the $\beta_I$ given by~\eqref{eq:beta_I_sol}, when substituted in~\eqref{eq:consistency_polynomial}, indeed satisfy the system of polynomial equations $P_I=0$, and that the choice $c_1=\beta_1^{-1}$ made there corresponds to a particular choice of the cosmological constant $\Lambda$ fixed by~\eqref{eq:beta_1_sol}. 

Note that a similar method as above works to obtain a solution for $c_I$ as well, starting from \eqref{eq:cosmological_constant} for $I=1$ to solve for $k^3$, and then solving for $c_{I>1}$ from a depressed cubic, which has a closed, but involved, real solution. The definition of $k$ is then used to fix $c_1$. Finally, we consider the possibility of solving the consistency equations \eqref{eq:c_consistency} for the set $\Lambda_I$ or for the Planck masses $m_I$, amounting to parametrising these in terms of the $c_I$ and the effective cosmological constant $\Lambda$ now taken as free parameters. We present these solutions for completeness, 
\begin{align}
\label{eq:Lambda_I-soln}
    \Lambda_I = \frac{1}{c_I^2}\left( \Lambda -m^4k^3 \frac{\beta_I}{m_I^2 c^{\;}_I}\right)\,,\quad \mathrm{or}\,, \quad
     m_I^2=m^4k^3\frac{\beta_I}{c_I(\Lambda-\Lambda_I c_I^2)  }\,, 
    \quad  I=1,\ldots,\mathcal{N} .
\end{align}

\section{Analysis of mass spectrum}
\label{sec:analysis_of_mass_spectrum}

Having established the existence of the proportional vielbein solutions, we now consider perturbations around these backgrounds and obtain the mass spectrum. Although it is not possible to disentangle the massive and massless modes for generic background solutions, we will show that this is possible for proportional backgrounds. By diagonalising the quadratic action, we obtain the familiar Fierz-Pauli mass terms for one massless and $\mathcal{N}-1$ massive spin-2 modes. This section concludes with an analysis of the positivity and structure of the eigenmasses.

\subsection{Quadratic action and the mass matrix}
\label{sec:quadratic_action}

In section~\ref{sec:perturbations}, we developed an all-order expansion of the action around generic background solutions. We now restrict this to quadratic order perturbations around proportional backgrounds. By using $\overline{e}^A_{I \; \mu}= c^{\;}_I \overline{e}^A_{\; \mu}$ and $g^I_{\mu \nu}= c_I^2 \overline{g}_{\mu \nu}$ in ~\eqref{eq:all_ord_exp} and~\eqref{eq:pot_expans}, and setting $\epsilon_I= \frac{1}{m_Ic_I}$ for canonical normalisation, we can expand the interaction term to quadratic order,
\begin{align}
   V=-2m^4\det(\sum_I \beta_I e_I) = V^{(0)}+ V^{(1)} + V^{(2)} + \Ord(m_I^{-3})\,,
\end{align}
where $V^{(0)}=-2m^4k^4 \sqrt{-\overline{g}}$, $\,V^{(1)}=-m^4 k^3\sqrt{-\overline{g}}\sum_{I = 1}^{\mathcal{N}}\frac{\beta_I }{m_I} h^\mu_{I \; \mu}$, and,
\begin{align}
   V^{(2)} = -2m^4k^2 \sqrt{-\overline{g}}\Bigg[ \sum_{I,J=1}^\mathcal{N}\frac{\beta_I \beta_J}{8 m_I m_J}\bigg \lbrace h^\mu_{I\; \mu}h^\nu_{J\; \nu}-h^\mu_{I\; \nu}h^\nu_{J\; \mu}-16\omega^\mu_{I \; \nu}\omega^\nu_{J \; \mu}  \bigg \rbrace\quad\notag\\ 
\label{eq:quad_pot}
   -k\sum_{I=1}^{\mathcal{N}}\frac{\beta_I}{8m_I^2c_I} 
   \bigg \lbrace h^\mu_{I \; \nu}h^{\nu}_{I \; \mu}-16\omega^\mu_{I \; \nu}\omega^\nu_{I \; \mu}\bigg \rbrace
   \Bigg]\,.
\end{align}
Note that $h^\mu_{I \; \nu}= \overline{g}^{\mu \lambda}_Ih^I_{\lambda \nu}=c_I^{-2}\overline{g}^{\mu \lambda}h^I_{\lambda \nu}$, and similarly for $\omega^I_{ \mu \nu}= c_I^2 \overline{g}_{\mu \lambda} \omega^\lambda_{I \; \nu}$.

Since the Einstein-Hilbert action is invariant under local Lorentz transformations, the $\omega^\mu_{I\; \nu}$ appear only in the interaction term. Their equations of motion can be obtained by either varying $V^{(2)}$ with respect to $\omega^I_{\mu \nu}$ or by inserting the Ansatz $\overline{e}^A_{I \; \mu}=c_I \overline{e}^A_{\; \mu}$ into~\eqref{eq:lin_LC}, which both yield,
\begin{align}
    \frac{\omega^\mu_{I\;\nu}}{m_I^{\;}c_I}=\frac{1}{k} \sum_{J=1}^\mathcal{N}\beta_Jc_J \frac{\omega^\mu_{J\;\nu}}{m_J^{\;}c_J}, \qquad  I=1,\ldots,\mathcal{N}.
\end{align}
Since the right hand side is independent of $I$, the solution is $\frac{\omega_{I}}{m_I^{\;} c_I}\equiv\epsilon_I\omega_I=\omega$, for all $I$ and for an undetermined 
$\omega^\mu_{~\nu}$. This corresponds to an overall Lorentz transformation of all vielbeins
which is a symmetry of the theory. As a result, all the $\omega$-dependent terms cancel in~\eqref{eq:quad_pot}. Then, together with the quadratic Einstein-Hilbert action,~\eqref{eq:quad_pot} yields the quadratic order multivielbein action in terms of metric perturbations alone,
\begin{align}
\label{eq:quadratic_action}
    \mathcal{S}^{(2)} = \int \text{d}^4x\sqrt{-\overline{g}}\sum_{I=1}^{\mathcal{N}}\bigg[\tfrac{1}{2}h_I^{\mu \nu}\mathcal{E}_{\mu \nu}^{\;\;\;\alpha \beta}h^I_{\alpha \beta}+\tfrac{\Lambda}{2}h_I^{\mu \nu}\para{\delta^\alpha_\mu \delta^\beta_\nu-\tfrac{1}{2}\overline{g}_{\mu \nu}\overline{g}^{\alpha \beta} }&h^{I}_{\alpha \beta}\notag\\
    \;-\tfrac{1}{4}\para{\delta^{\alpha }_\mu \delta^\beta_\nu -\overline{g}_{\mu \nu}\overline{g}^{\alpha \beta}}\sum_{J=1}^{\mathcal{N}} h_I^{\mu \nu}\mathcal{M}^{\;}_{IJ}&h^{J}_{\alpha \beta}
    \bigg].
\end{align}
Here $\mathcal{E}_{\mu \nu}^{\;\;\;\; \alpha \beta}$ is the differential operator given by~\eqref{eq:Lichnerowicz}. The background curvature terms of the linearised Einstein-Hilbert action have been eliminated for $\Lambda$, defined by~\eqref{eq:cosmological_constant}, which appears in the background equation \eqref{eq:background_solution} as the effective cosmological constant. The index structure of the second line is precisely of the form of the Fierz-Pauli mass term \cite{Fierz:1939ix,Pauli:1939xp}, but with a non-diagonal mass matrix  $\mathcal{M}_{I J}$ given by,
\begin{align}
\label{eq:Mass_matrix}
    \mathcal{M}_{IJ} = m^4 k^2 
        \begin{pmatrix} 
        \frac{\beta_1}{m_1^2c_1} \left( k - \beta_1 c^{\;}_1 \right) & -\frac{\beta_1 \beta_2}{m_1 m_2} & \ldots & -\frac{\beta_1 \beta_\mathcal{N}}{m_1 m_\mathcal{N}} \\
        -\frac{\beta_2 \beta_1}{m_2 m_1} & \frac{\beta_2}{m_2^2c^{\;}_2} \left( k - \beta_2 c_2 \right) & \ldots & -\frac{\beta_2 \beta_\mathcal{N}}{m_2 m_\mathcal{N}} \\
        \vdots & \vdots & \ddots & \vdots \\
        -\frac{\beta_\mathcal{N} \beta_1}{m_\mathcal{N} m_1} & -\frac{\beta_\mathcal{N} \beta_2}{m_\mathcal{N} m_2} & \ldots & \frac{\beta_\mathcal{N}}{m_\mathcal{N}^2c_\mathcal{N}} \left( k - \beta_\mathcal{N} c^{\;}_\mathcal{N} \right)
    \end{pmatrix}.
\end{align}
The eigenvalues are the squared Fierz-Pauli masses of the corresponding spin-2 eigensfields. Since the matrix $\mathcal{M}$ is symmetric, its eigenvalues are real. In Section~\ref{sec:mass_spectrum}, we will show that they are also non-negative, thus guaranteeing the absence of tachyonic modes. 

\subsection{Mass eigenfields}
\label{sec:mass_eigenstates}

The mass matrix arising from the quadratic action~\eqref{eq:Mass_matrix} can be written as the difference of a diagonal matrix $\mathcal{D}= \text{diag}(d_1,d_2,\ldots\,,d_\mathcal{N})$ with elements $d_I = m^4k^3 \frac{\beta_I}{m_I^2c^{\;}_I}$, and a rank-1 matrix $\textbf{v}\textbf{v}^{\T}$, where $\textbf{v}=m^2 k(\beta_1/m_1,\ldots\,,\beta_\mathcal{N}/m_\mathcal{N})^{\T}$, such that,
\begin{align}
\label{eq:diagplusrank1}
    \mathcal{M}=\mathcal{D}-\textbf{v}\textbf{v}^{\T}.
\end{align}
Since the parameters are not all independent, we may also choose to impose the parametric solutions \eqref{eq:beta_I_sol} or \eqref{eq:Lambda_I-soln} where, using the former, we obtain $d_I = \Lambda-\Lambda_I c_I^2$ and ${\textbf{v}=(mk)^{-2}(m_1c_1d_1,\ldots,m_{\mathcal{N}}c_{\mathcal{N}}d_{\mathcal{N}})^{\T}}$. The following analysis will be kept general to allow all solutions to the consistency equations \eqref{eq:c_consistency}.

The eigenvalues of matrices of the form~\eqref{eq:diagplusrank1} are well studied and do not exhibit a closed form in general \cite{Golub:1973,Bunch:1978, Thompson:1976}. The next subsection will provide a self-contained account of their structure. However, \eqref{eq:Mass_matrix} allows for the eigenvectors to be derived without knowing the explicit form of the eigenvalues. This is achieved by assuming that the eigenvalues $\mu_J^2$ of $\mathcal{M}$ are given,\footnote{Note that $\mu_I$ does not correspond to the mass of $h^I_{\mu \nu}$ in terms of which the mass matrix is not diagonal. Rather, the index $I=1,\ldots,\mathcal{N}$ on $\mu_I$ simply enumerates the eigenvalues of $\mathcal{M}$. To limit the proliferation of symbols, we do not introduce a new index for this purpose, but we will clarify the distinction when needed.} and satisfy $\mathcal{M} \textbf{x}_J = \mu_J^2 \textbf{x}^{\;}_J$ where the eigenvectors $\textbf{x}_J$ are to be determined. For each $J$, this results in a set of $\mathcal{N}$ homogeneous equations for the $\mathcal{N}$ components of $\mathbf{x}_J$ that we denote by $x_J^1, \ldots\,, x_J^{\mathcal{N}}$. These can be explicitly solved for in a manner analogous to standard eigenvector problems with known eigenvalues. Apart from an overall scaling of $\textbf{x}_J$, one finds that the components $\textbf{x}_J = (x_J^1, \ldots\,, x_J^\mathcal{N})^{\T}$ must have the form,
\begin{align}
    x^I_{J} =\frac{m_I c_I}{1-\frac{\mu_J^2}{d_I}}\,.
\end{align}
Using the normalised eigenvectors $\hat{\textbf{x}}_J=\frac{\textbf{x}_J}{|\textbf{x}_J|}$, we can construct the diagonalising matrix ${S= (\hat{\textbf{x}}_1,\ldots,\hat{\textbf{x}}_\mathcal{N})}$ with explicit components, 
\begin{align}
    S^{\;}_{IJ} = \frac{1}{|\textbf{x}_J|}\frac{m_I c_I}{1-\frac{\mu_J^2}{d_I}}, \qquad |\textbf{x}_J|= \sqrt{\sum_I \tfrac{m_I^2 c_I^2}{\big(1-\tfrac{\mu^2_J}{d_I}\big)^2}}\;.
\end{align}
Since $\mathcal{M}$ is symmetric, $S$ is an orthogonal matrix,\footnote{If two or more eigenvalues are the same, we can always use the Gram-Schmidt process to make the corresponding eigenvectors orthogonal.} thus $S^{-1}_{IJ}=S^{\T}_{IJ}=S^{\;}_{JI}$, and as expected $S^{\T}\mathcal{M}S= \text{diag}(\mu_1^2,\ldots\,,\mu^2_{\mathcal{N}})$ is diagonal. The fields in terms of which the mass matrix is diagonal are given by,
\begin{align}
\label{eq:eigenstates}
    f^I_{\mu \nu} = \sum_{J = 1}^\mathcal{N} S^{\T}_{IJ}h^J_{\mu \nu} = \frac{1}{|\textbf{x}_I|} \sum_{J = 1}^\mathcal{N} \frac{m_J c_J}{1-\frac{\mu^2_I}{d_J}}h^J_{\mu \nu}\,,
\end{align}
Then, substituting $h_I^{\mu \nu}=\sum_J S^{\;}_{IJ}f_J^{\mu\nu}$ into the quadratic action \eqref{eq:quadratic_action}, gives the diagonalised form,
\begin{align}
\label{eq:diagonalised_action}
    \mathcal{S}^{(2)} = \int \text{d}^4x\sqrt{-\overline{g}}\sum_{I = 1}^\mathcal{N} \bigg[
    \tfrac{1}{2}f_I^{\mu \nu}\mathcal{E}_{\mu \nu}^{\;\;\;\; \alpha \beta}f^I_{\alpha \beta}+\tfrac{\Lambda}{2}f^{\mu \nu}_I&\para{\delta^\alpha_\mu \delta^\beta_\nu-\tfrac{1}{2}\overline{g}_{\mu \nu}\overline{g}^{\alpha \beta}}f^I_{\alpha \beta}\notag\\
    \;-\tfrac{\mu_I^2}{4}f_I^{\mu\nu}&\para{\delta^{\alpha }_\mu \delta^\beta_\nu\, -\,\overline{g}_{\mu \nu}\overline{g}^{\alpha \beta}\,}f_{\alpha \beta}^I
    \,\bigg]\,,
\end{align}
which is a sum of Fierz-Pauli actions for spin-2 fields $f^I_{\mu\nu}$ with masses $\mu^{\;}_I$ on a background given by $\overline{g}_{\mu \nu}$. 
This form allows us to identify the multivielbein action as a theory of massive spin-2 fields. Note that if, instead, the background metric is taken to be, say, $\bar g_1=c_1^2\bar g$, then after readjusting the canonical normalizations, the new masses will scale to $\mu_I^2/c_1^2$ and the new cosmological constant becomes $\Lambda/c_1^2$.

\subsection{Mass eigenvalues}
\label{sec:mass_spectrum}

In this subsection, we investigate the properties of the eigenvalues of the mass matrix $\mathcal{M}$ and show that the spectrum contains one massless mode and $\mathcal{N}-1$ non-tachyonic massive modes. Instead of directly writing the characteristic polynomial for the eigenvalues, we develop the characteristic equation in a form that clearly brings out the properties of the eigenmasses. 

\subsubsection{Existence of non-tachyonic mass spectrum}
\label{sec:positive_roots}

We now consider the eigenvalues of a matrix of the form $\mathcal{M}= \mathcal{D}-\textbf{v}\textbf{v}^{\T}$. For eigenmasses $\mu$, the characteristic equation $p(\mu^2)=\det(\mathcal{M}-\mu^2\id)=0$ becomes,\footnote{It can easily be verified that $\mu^2=d_I$ is not a solution to the characteristic equation if $\beta_I\neq 0$ and all $d_I$ are distinct, thus guaranteeing the existence of $(\mathcal{D}-\mu^2\id)^{-1}$. The case where $d_I$ are not distinct will be discussed later.}
\begin{align}
    p(\mu^2)=\det(\mathcal{D}-\mu^2 \id -\textbf{v}\textbf{v}^{\T})= \det(\mathcal{D}-\mu^2 \id)\det(1-(\mathcal{D}-\mu^2 \id)^{-1}\textbf{v}\textbf{v}^{\T})=0\, .  
\end{align}
Since $(\mathcal{D}-\mu^2 \id)^{-1}\textbf{v}\textbf{v}^{\T}$ is a matrix of rank 1, the second determinant above takes the form $\det(\id-(\mathcal{D}-\mu^2 \id)^{-1}\textbf{v}\textbf{v}^{\T})=1-\textbf{v}^{\T}(\mathcal{D}-\mu^2 \id)^{-1}\textbf{v}$. Furthermore, since $\mathcal{D}-\mu^2 \id$ is diagonal, computing its determinant and inverse is trivial. Hence, 
\begin{align}
\label{eq:characteristic_eq}
    p(\mu^2)= \prod_{I=1}^{\mathcal{N}}\para{d_I-\mu^2}\hak{1-m^4k^2\sum_{J=1}^{\mathcal{N}}\frac{\beta^2_J }{m_J^2(d_J-\mu^2)}}=0\,.
\end{align}
Using the definition of $d_J = m^4k^3\frac{\beta_J}{m_J^2c_J^{\;}}$ to obtain $\frac{\beta^2_J}{m_J^2} = \beta_J c_J\frac{d_J}{m^4 k^3}$, and since $k = \sum_J \beta_J c_J$, the last bracket in \eqref{eq:characteristic_eq} simplifies to,
\begin{align}
\label{eq:secular_eq}
    \frac{1}{k}\sum_{I=1}^{\mathcal{N}} \beta_I c_I \left(1-\frac{d_I}{d_I-\mu^2}\right)= -\frac{\mu^2}{k}\sum_{I=1}^{\mathcal{N}}\frac{\beta_I c_I}{d_I-\mu^2}\,.
\end{align}
Putting these together gives the characteristic equation,
\begin{align}
\label{eq:characteristic_eq_2}
    p(\mu^2)= -\frac{\mu^2}{k}
    \sum_{I=1}^{\mathcal{N}}{\beta_I c_I}\prod_{\substack{J\neq I\\J=1}}^\mathcal{N}\para{d_J-\mu^2}
    =0\,.
\end{align}
Some immediate conclusions can be drawn from this equation: 
\begin{enumerate}[label =(\roman*)]
    \item $\mu^2 =0$ is a solution for all parameter values, thus always providing a massless spin-2 field in the spectrum which, from \eqref{eq:eigenstates}, is given by,
    \begin{align}
    \label{eq:massless_state}
        f^{(\mu^2=0)}_{\mu \nu}= \frac{1}{\sqrt{\sum_I m_I^2c_I^2}}\sum_{J=1}^\mathcal{N} m^{\;}_J c^{\;}_J h^J_{\mu \nu}\,.
    \end{align}
    \item We saw earlier that there always exist solutions to the consistency conditions~\eqref{eq:c_consistency} such that $\beta_I c_I>0 $, which in turn implies that $ k>0$ and $d_I= m^4k^3\frac{\beta_I c_I}{m_I^2c_I^2}>0$. Then, it is evident that for negative $\mu^2 = -|\mu^2|$, the sum in~\eqref{eq:characteristic_eq_2} is strictly positive,  
    \begin{align}
        \sum_{I=1}^{\mathcal{N}}{\beta_I c_I}\prod_{\substack{J\neq I\\J=1}}^\mathcal{N}\para{d_J+|\mu|^2}>0\, ,
    \end{align} 
    \item If some number $n$ of the $d_I$ have the same value $d$, then all terms in \eqref{eq:characteristic_eq_2} will have a common factor of $(d-\mu^2)^{n-1}$ and the spectrum contains $n-1$ degenerate states of mass $\mu^2=d$. The converse is also true in the sense that if $\mu^2=d_I$ is a root, then from direct substitution in \eqref{eq:characteristic_eq_2}, it follows that at least two of the $d$'s must be equal. 
\end{enumerate}

\subsubsection{Eigenmass structure and bounds}
\label{sec:eigenmasses}

We will now describe the general structure of the eigenmasses. Without loss of generality, we can relabel and rearrange the indices $I$ on the vielbeins such that the diagonal elements of $\mathcal{D}$ fulfil $d_I \leq d_{I+1}, \; I=1,\ldots\,,\mathcal{N}-1$.\footnote{Recall that while $d_I$ is connected to the perturbation $h^I_{\mu \nu}$, the labelling of the eigenvalues as $\mu_I^2$ is independent of $h^I_{\mu \nu}$ and is a matter of choice. In this subsection, we will make a convenient choice that correlates the label on $\mu^2_I$ to the $d_I$.} The non-zero eigenmasses are given by the roots of the sum in the characteristic equation \eqref{eq:characteristic_eq_2},
\begin{align}
\label{eq:characteristic_eq_3}
    q(\mu^2)=\sum_{I=1}^{\mathcal{N}}{\beta_I c_I}\prod_{\substack{J\neq I\\J=1}}^\mathcal{N}\para{d_J-\mu^2}=0\,.
\end{align}
Clearly, this is of order $\mathcal{N} -1$ in $\mu^2$ and for larger values of $\mathcal{N}$ explicit expressions for the mass eigenvalues do not exist. However, we can determine the arrangement of the eigenmasses using the intermediate value theorem, according to which, if the sign of $q(\mu^2)$ changes between two values of its argument, $\mu^2$, then the function must have a root between those values.

First, we assume that all $d_I$ are distinct and positive, such that $0<d_I<d_{I+1}$. Note that 
the $I^{\text{th}}$ term in the sum contains a product of factors $(d_J-\mu^2)$ for all $J\neq I$. Hence, only the $I^{\text{th}}$ term in the sum \eqref{eq:characteristic_eq_3} contributes to the value of $q(\mu^2=d_I)$, since every other summand contains the multiplicative factor $(d_I-\mu^2)$. We have shown that the smallest root of \eqref{eq:characteristic_eq_2} is $\mu^2=\mu^2_1=0$ at which $q(\mu^2=0)>0$. In the interval $0<\mu^2\leq d_1$, $q(\mu^2)$ remains positive and has no new roots. In particular, $q(\mu^2=d_1)>0$ since then the $I=1$ term is positive and all other terms vanish.

As $\mu^2$ becomes larger than $d_1$, all terms in \eqref{eq:characteristic_eq_3} change sign except for the $I=1$ term which is the only one not containing the factor $(d_1-\mu^2)<0$. This only positive term finally vanishes for $\mu^2=d_2$ ensuring that $q(\mu^2=d_2)<0$. Hence, $q(\mu^2)$ changes sign in the interval $(d_1,d_2)$, so there exists a root $d_1<\mu_2^2<d_2$. Increasing $\mu^2$ beyond $d_2$, it is easily seen that $q(\mu^2=d_3)>0$ which establishes the existence of a new root $d_2<\mu_3^2<d_3$. Proceeding similarly, one finds new roots for every interval $d_{I-1}<\mu_I^2<d_I$. 

We have also seen earlier, at the end of the previous subsection, that if and only if $d_{I-1}=d_I$, then there exists a root $\mu_I^2=d_I$. Hence, relaxing the condition on the strict ordering of the $d_I$ to $d_I \leq d_{I+1}, \; I=1,\ldots\,,\mathcal{N}-1$, the eigenvalues fulfil the bounds,
\begin{align}
\label{eq:root_sandwich}
    0=\mu_1^2 < d_1 \leq \mu_2^2 \leq d_2 \leq\ldots \leq d_{\mathcal{N}-1}\leq \mu^2_\mathcal{N}\leq d_\mathcal{N}\,.
\end{align}
Thus, all the non-zero eigenvalues are sandwiched between the matrix elements of $\mathcal{D}$. In particular, given that we have chosen the solutions to the consistency conditions such that $d_I >0$ (as described in Section~\ref{sec:existence_of_solutions}), the eigenmasses are always bounded from below by $0$ and from above by $d_\mathcal{N}$. As an illustration, the $\mathcal{N}=2$ and $\mathcal{N}=3$ cases are discussed explicitly below.

In \eqref{eq:root_sandwich}, the masses are bounded by the $d_I= m^4k^3\beta_I/m_I^2c_I$. These are complicated functions of the free parameters $\beta_I, m_I$, and $\Lambda_I$ due to the dependence of the $c_I$ on these parameters, as determined by the conditions \eqref{eq:c_consistency}. However, this simplifies if we use the alternative parametrisations of Section~\ref{sec:parametric_sol}. In the parametrisation where $\Lambda_I$ are eliminated in favour of the $c_I$ through equation \eqref{eq:Lambda_I-soln}, the expressions for the bounds will remain unchanged, except that now the $c_I$ are treated as independent theory parameters along with $\beta_I$ and $m_I$.  

Alternatively, $\beta_I$ can be traded for the new free parameters $c_I$ through equation \eqref{eq:beta_I_sol}. In this setup, $d_I=\Lambda - \Lambda_I c_I^2$ and the mass bounds are given explicitly in terms of the free parameters,
\begin{align}
\label{eq:tuning_sandwich}
    0=\mu^2_1<\Lambda - \Lambda_1^{\;} c_1^2 \leq \mu^2_2\leq \Lambda- \Lambda^{\;}_2 c_2^2  \leq \ldots \leq \mu^2_\mathcal{N}\leq \Lambda - \Lambda^{\;}_\mathcal{N}c_\mathcal{N}^2\, .
\end{align}
The conditions $\Lambda-\Lambda^{\;}_Ic_I^2>0$ for all $I$, correspond to $\beta_Ic_I>0$, as implied by \eqref{eq:beta_I_sol}, which guarantee that zero is the smallest eigenvalue. From this form, it is easy to see that:
\begin{enumerate}[label=(\roman*)]
    \item If $\Lambda_I =0$ for all $I$, then all $\mathcal{N}-1$ non-zero eigenmasses are degenerate, $\mu^2_I = \Lambda$. More generally, if $n$ of the $\Lambda_I$ vanish, then $n-1$ of the corresponding masses are degenerate and equal to $\Lambda$.
    \item In flat spacetime ($\Lambda=0$) the masses are non-tachyonic only if $\Lambda_I < 0$. The AdS case ($\Lambda <0$) is discussed below.
\end{enumerate}

\noindent In the new parametrisation, the characteristic equation $p(\mu^2)=0$ depends only on the free parameters. Discarding an overall factor, the part from which the mass eigenvalues can be calculated is,
\begin{align}
\label{eq:secular_equation_2}
    \Tilde{p}(\mu^2)=\mu^2\sum_{I=1}^{\mathcal{N}}m_I^2c_I^2 \prod_{\substack{J\neq I\\J=1}}^\mathcal{N}\para{1 - 
    \frac{\mu^2}{\Lambda - \Lambda_J^{\;} c_J^2} }.
\end{align}
The parametrisation in terms of a fixed cosmological constant is especially useful when considering perturbations around de Sitter (dS) and Anti-de Sitter (AdS) background solutions, where the Higuchi~\cite{Higuchi:1986py} and Breitenlohner-Freedman~\cite{Breitenlohner:1982jf} bounds restrict physical masses on dS and AdS respectively. 

The bounds \eqref{eq:tuning_sandwich} imply that dS backgrounds $(\Lambda>0)$ with non-tachyonic masses are guaranteed to exist for positive or negative $\Lambda_I$, as long as $\Lambda-\Lambda_I^{\;} c_I^2>0$.\footnote{The reindexing is such that $d_I \leq d_{I+1}$, which implies $\Lambda^{\;}_\mathcal{N}c_\mathcal{N}^2\leq \ldots \leq \Lambda^{\;}_2c_2^2 \leq \Lambda^{\;}_1c_1^2$.}  However, the smallest non-zero eigenmass $\mu_2^2$, must be greater than the Higuchi mass, $\mu^2_2 > m_{\text{H}}^2=\frac{2}{3}\Lambda$.\footnote{The bound $\mu^2_2 \geq m_{\text{H}}^2$ is necessary to avoid the Higuchi ghost. But we exclude $\mu_2^2 = m_{\text{H}}^2$, for which the quadratic theory develops a partial massless symmetry, resulting in the $\mu_2$-mode having only 4 propagating helicities. Such a symmetry does not survive beyond quadratic order \cite{Boulanger:2024hrb,Apolo:2016vkn,Joung:2014aba}, implying that, in such a setup, the 0-helicity mode is strongly coupled.} This implies restrictions on $\Lambda^{\;}_1c^2_1$ and $\Lambda^{\;}_2c^2_2$, e.g. the values $\Lambda^{\;}_1c^2_1=\Lambda^{\;}_2 c^2_2=\tfrac{1}{3}\Lambda$ would force $\mu^2_2=m_{\text{H}}^2$ which is excluded.  

In AdS backgrounds ($\Lambda<0$), negative squared masses are allowed without causing instabilities; thus, the smallest eigenmass is not necessarily zero. To obtain such solutions, one must go outside the region where we have proved that real solutions to the consistency conditions always exist. However, numerical calculations suggest they do exist. Nevertheless, in such a case, the masses must still be bounded by the Breitenlohner-Freedman mass $\mu_1^2 > m_{\text{BF}}^2= \frac{3}{4}\Lambda$. To obtain a bound on the free parameters, we note that the smallest eigenvalue of $\mathcal{M}= \mathcal{D}-\textbf{v}\textbf{v}^{\T}$ is bounded from below by $d_1-|\textbf{v}|^2$~\cite{Golub:1973}. This can be used to obtain a sufficient, although not necessary, condition,
\begin{align}
\label{eq:AdS_mass_bound}
     \frac{\Lambda}{4}> \Lambda^{\;}_1c_1^2 +\frac{1}{m^4k^4}\sum_{I=1}^\mathcal{N} m_I^2c_I^2(\Lambda-\Lambda^{\;}_Ic_I^2)^2\qquad (\text{AdS})\,.
\end{align}
It is worth emphasising that we are not forced to have negative $\mu_I^2$ on AdS, only that they are not problematic if the above bound is respected.

Consequently, although the theory potentially accommodates parameters that violate the Higuchi and Breitenlohner-Freedman bounds, a wide set of parameters exists within the consistent region of the parameter space for all $\Lambda$. In conclusion, the analysis of the eigenmass structure and the derived bounds for the mass spectrum of the perturbations propagating on a proportional background have shown that the theory can be parameterised to ensure non-tachyonic mass eigenvalues, adhering to the related consistency conditions and physical constraints.

\subsubsection{Geometric interpretation of the characteristic equation}

We will now obtain an expression for the coefficients of the characteristic polynomial corresponding to  \eqref{eq:characteristic_eq_2} and also point out an interesting geometric interpretation of the eigenmass problem. The form that was useful for finding the bounds on eigenvalues is, up to an overall multiplicative factor, 
\begin{align}
\label{eq:secular_equation_3}
    \Tilde{p}(x)=x\sum_{I=1}^{\mathcal{N}}a_I \prod_{\substack{J\neq I\\J=1}}^\mathcal{N}\para{1 - b_J\,{x}}=0\,,
\end{align}
where, for this section only, we use the notation $x=\mu^2$. This is equivalent to \eqref{eq:characteristic_eq_2} with 
$a_I=\beta_Ic_I/d_I$, $b_J=d_J^{-1}$ and recovers 
\eqref{eq:secular_equation_2} for $a_I=m_I^2c_I^2$, $b_J=(\Lambda-\Lambda_J^{\;} c_J^2)^{-1}$. 

Let us start with the diagonal matrix $\mathcal{B}=\mathcal{D}^{-1}= \text{diag}(b_1,b_2,\ldots\,,b_\mathcal{N})$ and consider $\det(\id -\mathcal{B}x)= \prod_{J=1}^{\mathcal{N}}(1- b_J x)$. On differentiating the product form, it is obvious that,
\begin{align}
\label{eq:characteristic-pol1}
    -\sum_{I=1}^{\mathcal{N}}a_I\frac{\partial}{\partial b_I}
    \Big[\det(\id -\mathcal{B}x)\Big]= 
    x\sum_{I=1}^{\mathcal{N}}a_I \prod_{\substack{J\neq I\\J=1}}^\mathcal{N}\para{1 - b_J\,{x}}=\Tilde{p}(x)\, .
\end{align}
Now, consider the $\mathcal{N}$-dimensional space spanned by the $b_I$. Setting $\det(\id - \mathcal{B}x)$ to be constant (i.e. $b_I$ independent), defines a one-parameter family of $(\mathcal{N}-1)$-dimensional hypersurfaces parametrised by $x \neq 0$. The left-hand side of \eqref{eq:characteristic-pol1} computes the projection of the normal to these hypersurfaces along a given $\mathcal{N}$-dimensional vector with components $a_I$. Thus, finding the non-zero roots of $p(x)=0$ corresponds to finding hypersurfaces to which the vector $a_I$ is tangent.

To obtain the coefficients of the characteristic polynomial, note that the determinant in the left-hand side of equation \eqref{eq:characteristic-pol1} can also be expanded in terms of the elementary symmetric polynomials $\sigma_n(\mathcal{B})$ given by \eqref{eq:el_sym_pol}, but which for a diagonal matrix $\mathcal{B}$ have very simple forms, ${\sigma_0=1}$, ${\sigma_1=\sum_I b_I}$, ${\sigma_2=\sum_{I<J} b_I b_J\,}$, $\ldots\,, \,\sigma_n=\sum_{I_1<I_2<\ldots<I_n} b_{I_1}b_{I_2}\ldots b_{I_n}$, and ${\sigma_{\mathcal{N}}= b_1 \ldots b_{\mathcal{N}}}$. Then, substituting ${\det(\id -\mathcal{B}x)=\sum_{n=0}^{\mathcal{N}}(-1)^n\sigma_n(\mathcal{B})\, x^n}$ in the left hand side of \eqref{eq:characteristic-pol1} immediately gives the expression for the characteristic polynomial as,
\begin{align}
\label{eq:characteristic-pol2}
    \Tilde{p}(x)=-\sum_{n=1}^{\mathcal{N}} (-1)^n
    \left(\sum_{I=1}^{\mathcal{N}}a_I\frac{\partial}{\partial b_I}\right)\sigma_n(\mathcal{B})\, x^{n}\,.
\end{align}
The coefficients of the various powers of $x$ can be easily evaluated since the terms in $\sigma_n$ are multilinear in the $b_I$.

\subsubsection{Some explicit solutions}
\label{sec:degenerate}

Given the $\mathcal{N}^{\text{th}}$-order nature of the characteristic equation~\eqref{eq:characteristic_eq_2}, it is generally unfeasible to determine the analytic form of the eigenvalues for $\mathcal{N}>5$. Additionally, the cases $\mathcal{N}=4,5$ do not yield significant insights. Therefore, we will focus on presenting the $\mathcal{N}=2$ and $\mathcal{N}=3$ cases in the following discussion. Recall that, as shown in Section~\ref{sec:positive_roots}, the massless spin-2 field always has the same form for all $\mathcal{N}$,
\begin{align}
    f^{(1)}_{\mu \nu}= \frac{1}{\sqrt{\sum_I m_I^2c_I^2}}\sum_{J=1}^\mathcal{N} m^{\;}_J c^{\;}_J h^J_{\mu \nu}\,,
\end{align}
showing that all the fields $h^J_{\mu \nu}$ (or the $e^A_{J\; \mu}$ at the non-linear level) contribute to the massless spin-2 mode. 

Gravitational waves are observed to be practically massless. Hence, if we assume that $g^{(1)}_{\mu\nu}$ is the gravitational metric that couples to the standard model matter, then its perturbation $h^{(1)}_{\mu\nu}$ must have a large overlap with $f^{(1)}_{\mu \nu}$. This requires $m_1 c_1 \gg m_J c_J$ for all $J\neq 1$ and is consistent with the large observed value of the physical Planck mass $M_{\text{Pl}} \sim m_1$.

For $\mathcal{N}=2$, there is only one non-zero eigenmass, and the characteristic equation~\eqref{eq:characteristic_eq_2} reduces to a linear equation. The solution in two different parametrisations is given by,
\begin{align}
    \mu_2^{2}&=m^4k^2\beta_1c_1 \beta_2c_2 \para{\frac{1}{m_1^2c_1^2}+\frac{1}{m_2^2c_2^2}}=d_1d_2\;
    \frac{m_1^2c_1^2+m_2^2c_2^2}
    {m_1^2c_1^2d_1+m_2^2c_2^2d_2}\; ,
\end{align}
where $d_I = m^4k^3 \frac{\beta_I}{m_I^2c_I}=\Lambda-\Lambda_Ic^2$. Using~\eqref{eq:eigenstates}, the corresponding massive field simplifies to,
\begin{align}
    f^{(2)}_{\mu \nu} = \frac{1}{\sqrt{m_1^2c_1^2+m_2^2c_2^2}}\hak{m_2c_2 h^{(1)}_{\mu \nu}-m_1c_1 h^{(2)}_{\mu \nu}}.
\end{align}
Clearly the diagonalisation is achieved by a rotation, 
\begin{align}
    \begin{pmatrix}
        f^{(1)}_{\mu \nu}\\
        f^{(2)}_{\mu \nu}
    \end{pmatrix}= \begin{pmatrix}
        \cos \theta & \sin \theta\\
        -\sin \theta & \cos \theta
    \end{pmatrix}\begin{pmatrix}
        h^{(1)}_{\mu \nu}\\
        h^{(2)}_{\mu \nu}
    \end{pmatrix}\,,
\end{align}
with mixing angle given by $\tan\theta=m_1 c_1/m_2 c_2$ in terms of the ratio of Planck masses. This is a special case of bimetric theory \cite{Hassan:2012wr}.

The eigenmasses for $\mathcal{N}=3$ are more involved as $q(\mu^2)$ is quadratic and the roots have a fairly complicated form in terms of the theory parameters,
\begin{align}
    \mu^2_{\pm}&=A\pm \sqrt{A^2-B}\,,\\
    A & = \frac{1}{2k}\Big [\beta_1 c_1 \para{d_2+d_3}+\beta_2 c_2 \para{d_1+d_3}+\beta_3 c_3 \para{d_2+d_1}\Big]\,,\notag\\
    B & = \frac{1}{k}\Big[\beta_1c_1d_2d_3+\beta_2c_2d_1d_3+\beta_3c_3d_1d_2 \Big], \notag
\end{align}
where $d_I = \beta_I \frac{m^4 k^3}{m_I^2c_I}=\Lambda-\Lambda_Ic_I^2$. The corresponding eigenfields are,
\begin{align}
    f^{(\pm)}_{\mu \nu} = \frac{
     \frac{m_1 c_1 d_1}{d_1 - \mu^2_{\pm}} h^{(1)}_{\mu \nu} 
    +\frac{m_2 c_2 d_2}{d_2 - \mu^2_{\pm}} h^{(2)}_{\mu \nu}
    +\frac{m_3 c_3 d_3}{d_3 - \mu^2_{\pm}} h^{(3)}_{\mu \nu}}
    {\sqrt{\frac{m_1^2 c_1^2 d_1^2}{(d_1 - \mu^2_{\pm})^2} + \frac{m_2^2 c_2^2 d_2^2}{(d_2 - \mu^2_{\pm})^2} + \frac{m_3^2 c_3^2 d_3^2}{(d_3 - \mu^2_{\pm})^2}}}\,.
\end{align}
Again, we see that all $h^{I}_{\mu \nu}$ contribute to the massive spin-2 modes.

As shown in Section~\ref{sec:positive_roots}, only if some of the $d_I$ are the same, we will get degenerate eigenmasses. As previously mentioned, such a situation arises, for example, if $\Lambda_I =0$ for all $I$ and results in degenerate masses which are equal to the effective cosmological constant,
\begin{align*}
    m_{\text{FP}}^2=\Lambda = m^8 \lambda^2 \hak{\sum_J \frac{\beta_J^2}{m_J^2}}^3\,,
\end{align*}
for a parameter $\lambda$ set by the choice of the solution.

While the massless spin-2 field $f^{(1)}_{\mu \nu}$ is unaffected, the degeneracy means that~\eqref{eq:eigenstates} does not produce different field configurations for the $\mu_I^2 = m_{\text{FP}}^2$. Thus one needs to manually construct orthonormal states in this subspace, e.g., by the Gram-Schmidt process. For $\mathcal{N}=3$, the degenerate eigenspace is spanned, for example, by the two fields,
\begin{align*}
    f^{(2)}_{\mu \nu}= \frac{\frac{\beta_2}{m_2}h^{(1)}_{\mu \nu} -\frac{\beta_1}{m_1}h^{(2)}_{\mu \nu}}{\sqrt{\frac{\beta_1^2}{m_1^2}+\frac{\beta_2^2}{m_2^2}}}, \qquad
    f^{(3)}_{\mu \nu}= \frac{\frac{\beta_3}{m_3}h^{(1)}_{\mu \nu} -\frac{\beta_1}{m_1}h^{(3)}_{\mu \nu}}{\sqrt{\frac{\beta_1^2}{m_1^2}+\frac{\beta_3^2}{m_3^2}}}\,,
\end{align*}
which can be orthogonalised if desired.

The above analysis explicitly shows that the massive and massless modes of the spin-2 fields are distributed over all the vielbeins in~\eqref{eq:MM_action}, but that there always are $5(\mathcal{N}-1)$ massive and $2$ massless propagating modes at a quadratic level. This agrees with the non-linear analysis performed in~\cite{Hassan:2018mcw}, showing that the theory can be interpreted as a theory of interacting spin-2 fields.

\section{Summary and discussion}
\label{sec:conclusion}

After a review of the multivielbein model and its equations of motion, we considered perturbations around generic background solutions. We first developed an all-order expansion of the vielbein in terms of the metric and Lorentz field perturbations applicable to any vielbein theory and then outlined the procedure for expanding the multivielbein action to any order. The equations of motion and Lorentz constraints were then expanded explicitly to linear order around generic background solutions. To study the mass spectrum, we considered a class of background solutions to the source-free non-linear equations where all $\mathcal{N}$ metrics are proportional to each other. These turned out to be vacuum Einstein spacetimes with an effective cosmological constant, around which perturbations have Fierz-Pauli-type masses. We showed that such classes of solutions always exist and discussed their existence conditions in different parameterisations of the theory. Expanding the action to quadratic order around proportional backgrounds, we identified the mass matrix $\mathcal{M}_{IJ}$ and showed that on diagonalisation, it leads to Fierz-Pauli terms for one massless and $\mathcal{N}-1$ massive perturbations. The massless mode is universal, and the corresponding eigenfield is found for all $\mathcal{N}$. For the non-zero eigenvalues, a closed form expression is not known for $\mathcal{N}>5$ and is anyway not practical beyond $\mathcal{N}=3$. However, due to the particular form of the mass matrix, we found lower and upper bounds on each mass \eqref{eq:tuning_sandwich} showing, in particular, that a non-tachyonic spectrum always exists, regardless of the sign of the cosmological constant. Some relationships and restrictions applicable to the masses and theory parameters were also discussed following \eqref{eq:tuning_sandwich}, including the known instabilities in dS and AdS spacetimes. Finally, we gave the explicit expressions for $\mathcal{N}=2,3$. 

The multivielbein theory can be explored further in different directions. Theoretically, one may attempt to understand the structure of the larger set of ghost-free multi-spin-2 theories beyond the model considered here. Already for two spin-2 fields, the known set of bimetric interactions is larger than the ones described by \eqref{eq:HSM_interaction} for $\mathcal{N}=2$. While some simple extensions of the present model can be constructed, the general structure of ghost-free multiple spin-2 interactions has yet to be discovered. Also, while the absence of ghosts in the model \eqref{eq:MM_action} was argued in \cite{Hassan:2018mcw}, the existence of higher-level constraints that remove the ghosts needs to be established explicitly. We hope to report on this in the near future. It is also interesting to find a formulation of the theory given entirely in terms of the metrics beyond the method used in \cite{Hassan:2012wt}. 

Another interesting direction is the investigation of the phenomenological implications of the model similar to what has been worked out for bimetric theory, see, for example, \cite{Luben:2018ekw} and references therein. 
More specifically, some interesting applications include 
interpreting the massive states as dark matter  \cite{Babichev:2016bxi}, or the implications for the expansion rate and the Hubble tension \cite{Luben:2019yyx,Dwivedi:2024okk}, to name a few. We hope to come back to some of these issues in the near future.

\acknowledgments
We want to thank Mikica Kocic and Marcus Högås for valuable discussions and suggestions. 

\newpage
\bibliography{biblio.bib}

\end{document}